\documentclass{iaus}
\usepackage{graphicx}
\usepackage{natbib}
\title[GMIMS] 
{GMIMS: The Global Magneto-Ionic Medium Survey}

\author[Wolleben et al.]   
{M. Wolleben$^{1,12}$ T. L. Landecker$^1$, E. Carretti$^2$, J. M. Dickey$^3$, \break A. Fletcher$^4$, B. M. Gaensler$^5$, J. L. Han$^6$, M. Haverkorn$^7$, \break J. P. Leahy$^8$, N. M. McClure-Griffiths$^9$, D. McConnell$^9$, W. Reich$^{10}$ \and A. R. Taylor$^{11}$}

\affiliation{$^1$ NRC Herzberg Institute of Astrophysics, DRAO, Penticton, BC, V2A6J9, Canada \break
$^2$ INAF – Istituto di Radioastronomia, Via Gobetti 101, 40129 Bologna, Italy \break
$^3$ Physics Department, University of Tasmania, Hobart, TAS 7001, Australia \break
$^4$ School of Mathematics and Statistics, Newcastle University, NE1 7RU, U.K.\break
$^5$ Institute of Astronomy, School of Physics, The University of Sydney, NSW 2006, Australia \break
$^6$ National Astronomical Observatories, Chinese Academy of Sciences, Beijing 100012, China \break
$^7$ Jansky Fellow, National Radio Astronomy Observatory, Charlottesville, VA 22903, USA \break
$^8$ School of Physics \& Astronomy, The University of Manchester, Cheshire SK11 9DL, UK \break
$^9$ Australia Telescope National Facility, CSIRO, PO Box 76, Epping, NSW 2121, Australia \break
$^{10}$ Max-Planck-Institut f\"ur Radioastronomie, Auf dem H\"ugel 69, 53121 Bonn, Germany \break
$^{11}$ Centre for Radio Astronomy, University of Calgary, Calgary, AB T2N 1N4, Canada \break
$^{12}$ Covington Fellow \break
}

\pubyear{2009}
\volume{259}  
\pagerange{100--100}
\date{"10th Dec 2008"  and in revised form ??}
 \jname{Cosmic Magnetic Fields: From Planets,
to Stars and Galaxies} \editors{K.G. Strassmeier, A.G. Kosovichev \&
J.E. Beckman, eds.}

\begin{document}

\maketitle

\begin{abstract}

The Global Magneto-Ionic Medium Survey (GMIMS) is a project to map the diffuse polarized emission over the entire sky, Northern and Southern hemispheres, from 300 MHz to 1.8 GHz. With an angular resolution of 30 - 60 arcmin and a frequency resolution of  1 MHz or better, GMIMS will provide the first spectro-polarimetric data set of the large-scale polarized emission over the entire sky, observed with single-dish telescopes. GMIMS will provide an invaluable resource for studies of the magneto-ionic medium of the Galaxy in the local disk, halo, and its transition. 


\end{abstract}

\firstsection

\section{Motivation}

Polarization of the diffuse synchrotron emission from the Galaxy provides a sensitive probe of the magnetic fields and warm and hot plasmas making up the Galactic magneto-ionic medium (MIM). Polarization maps at decimetre wavelengths thus reveal intriguing rotation measure structures that are otherwise undetectable. These structures contain information about the Galactic magnetic field and the distribution of the ionized gas. In short, the physics of the MIM is imprinted on the diffuse polarized emission. So far, single-frequency (or narrow-band) polarization surveys have only allowed rather qualitative studies due to ambiguities in the data. Wide-band polarization data, as they are now underway through various survey projects, can resolve these ambiguities and will allow quantitative analysis of the diffuse polarized emission of the Galaxy.

Figure 1 summarizes our current knowledge of the northern polarized sky at 408 MHz (left) and 1.4 GHz (right). The polarization vectors at 408 MHz were measured more than 30 years ago. Similar data at 1.4 GHz were only recently superseded by the new DRAO Low-Resolution Polarization Survey. Data with higher angular resolution have been observed by, for example, the EMLS \citep{2004mim..proc...45R} and CGPS (Landecker et al., in prep.) in the northern sky, as well as SGPS \citep{2006ApJS..167..230H} and S-PASS \citep{2008arXiv0806.0572C} on the southern sky. GMIMS aims at superseding and complementing these data through its frequency coverage (or, more importantly for rotation measure studies, through its $\lambda^2$-coverage).

\begin{figure}[t]
 \includegraphics[scale=0.35]{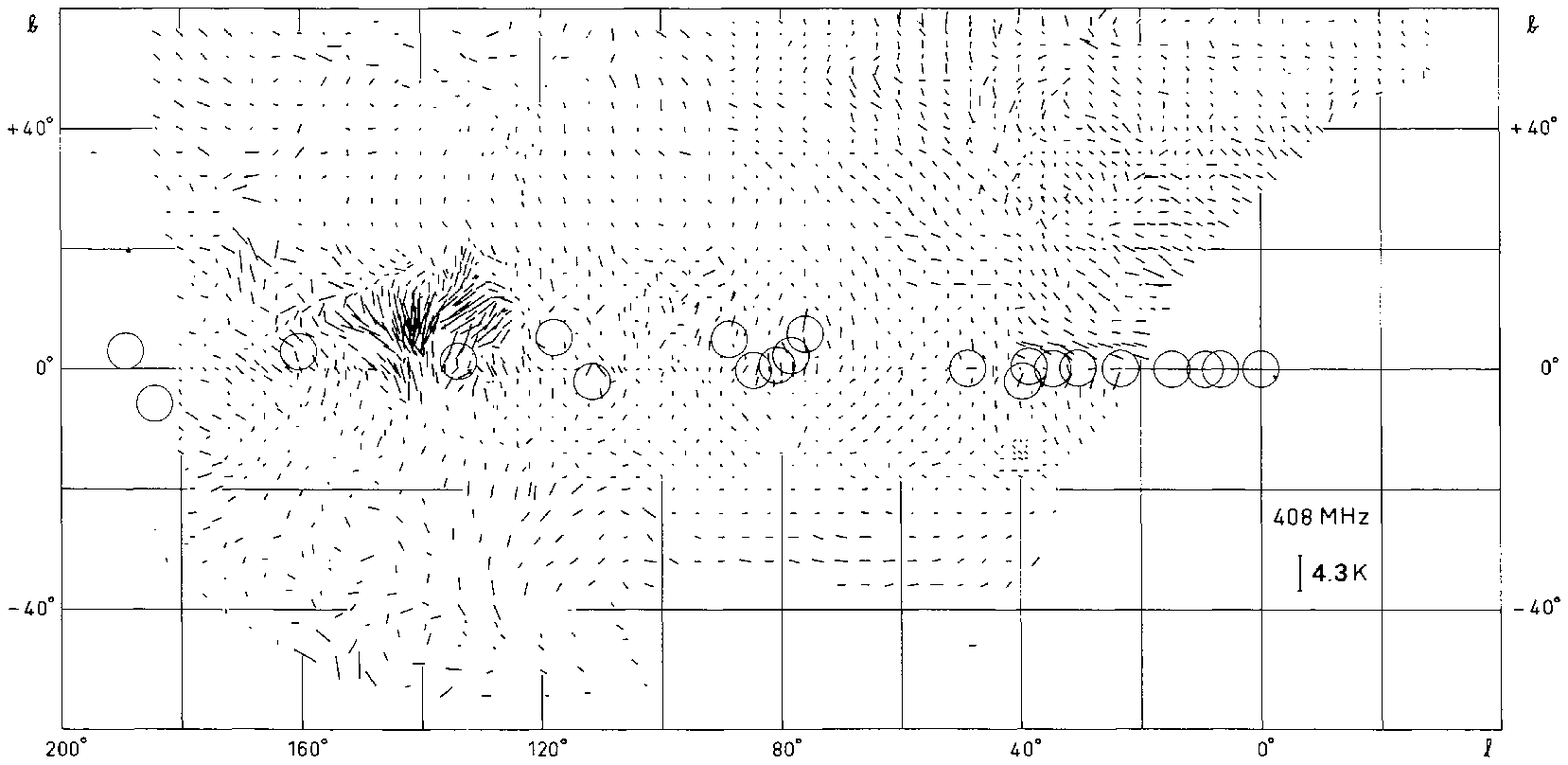}
 \includegraphics[scale=0.35]{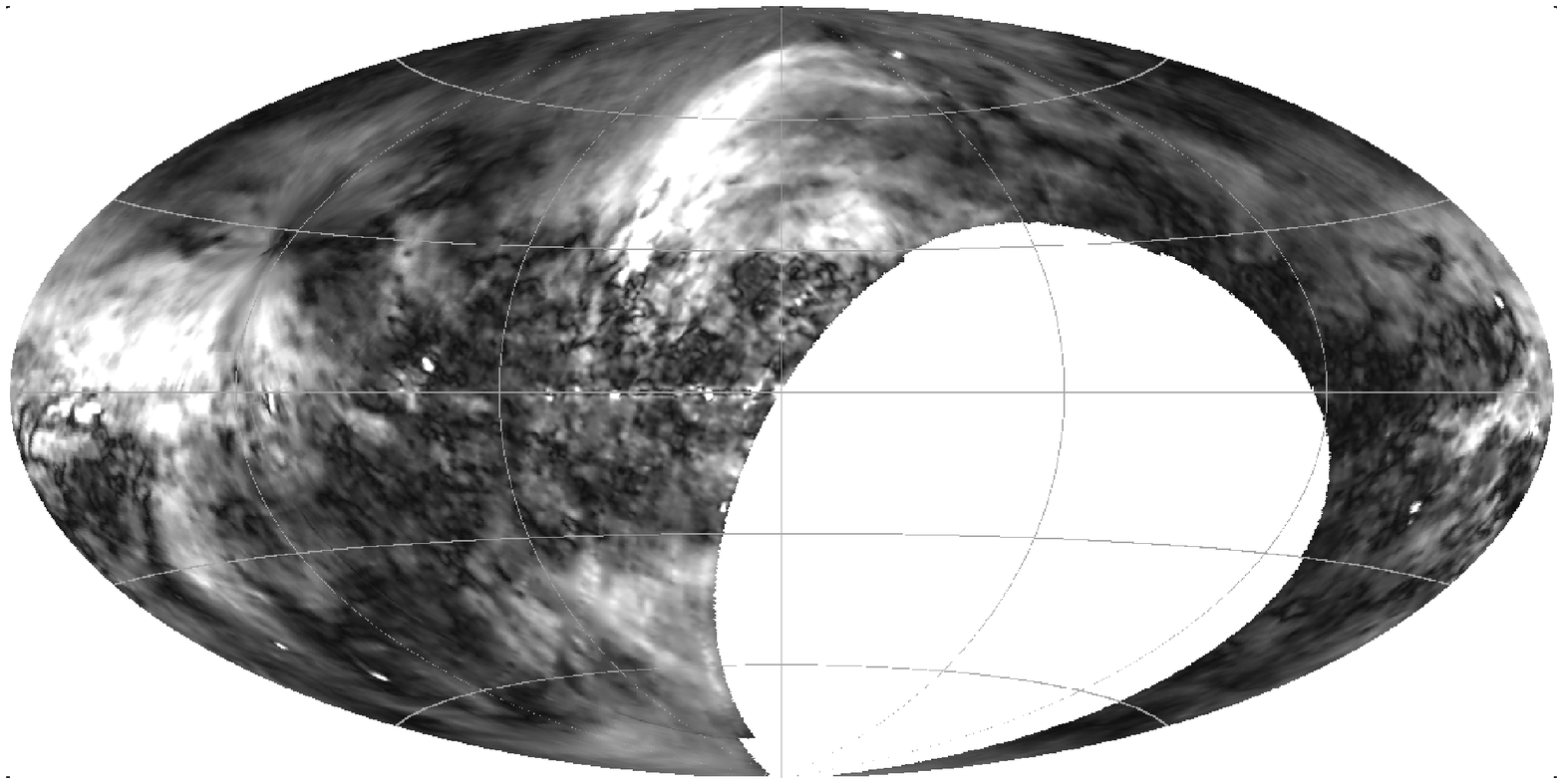}
  \caption{Polarization maps of the northern sky in Galactic coordinates at 408 MHz \citep[left, from][]{1976A&AS...26..129B} and 1.4 GHz \citep[right, from][]{2006A&A...448..411W}. See Reich \& Reich (this issue) for an all-sky version of the 1.4 GHz survey.}
\end{figure}

\section{Strategy}

The goal of GMIMS is to provide absolutely calibrated Stokes I, U, and Q data cubes of the entire sky. The intended frequency band (300 MHz to 1.8 GHz) has been split into Low-, Mid-, and High-Band. GMIMS thus comprises six {\it component surveys}, three in the North and three in the South. The first of these, the High-Band North (1.3 - 1.8 GHz), uses the DRAO 26-m Telescope and is now 25\% complete. Its southern counterpart is the STAPS project (PI M. Haverkorn), currently being carried out with the Parkes telescope. A very successful pilot study for the Low-Band South (300 - 900 MH) survey has recently been completed with the Parkes telescope. The other component surveys are in the planning stages.

\section{From 2-D to 3-D Polarimetry}

GMIMS is made possible by new developments and technologies in radio astronomy: new wide-band feeds, wide-band digital polarimeters, and rotation measure synthesis (RMS); all of which only recently became available to polarization science. The design of wide-band antennas benefits from modern EM-simulation software and is motivated by the SKA; field programmable gate arrays (FPGA) are now powerful enough to provide the basis for digital wide-band polarimeters; and \citet{2005A&A...441.1217B} demonstrated the potential of RMS in analyzing wide-band polarization data.

\end{document}